\documentclass[%
 aip,
 amsmath,amssymb,
 reprint,%
]{revtex4-1}

\usepackage{dcolumn}% Align table columns on decimal point
\usepackage{bm}% bold math
\usepackage[utf8]{inputenc}
\usepackage[T1]{fontenc}
\usepackage{mathptmx}
\usepackage{graphicx,color,epsf,bm}
\usepackage{amsmath,amsfonts,amssymb,amscd,bm}
\usepackage{soul}

\usepackage[colorlinks=true, urlcolor=blue]{hyperref}
\hypersetup{
	colorlinks=true,
	citecolor = blue,
	linkcolor=blue,
	filecolor=magenta,      
	urlcolor=cyan,
}

\newcommand\figref[1]{Fig.~\ref{#1}}

\newcommand\sectref[1]{Section~\ref{#1}}

\newcommand{\Omegarm}   {\mathrm{\Omega}}

\newcommand{\bfH}   {\mathbf{H}}
\newcommand{\bfB}   {\mathbf{B}}

\newcommand{\bfb}   {\mathbf{b}}

\newcommand{\bfe}   {\mathbf{e}}
\newcommand{\bfh}   {\mathbf{h}}
\newcommand{\bfj}   {\mathbf{j}}

\newcommand{\bft}   {\mathbf{t}}

\newcommand{\bfJ}   {\mathbf{J}}

\newcommand{\bfT}   {\mathbf{T}}

\newcommand{\mueff}   {\mu_{\mathrm{eff}}}
\newcommand{\Peff}   {P_{\mathrm{eff}}}

\newcommand{\opcurl}{\operatorname{curl}}

\begin{document}

%\preprint{AIP/000-XXX}
\title[Effective Medium Transformation]{Effective Medium Transformation:\\
 the Case of Stratified Magnetic Structures}
% Force line breaks with \\

\author{Markus Sch\"obinger}
\email{markus.schoebinger@tuwien.ac.at}
 \affiliation{Technische Universit\"at Wien, Institute for Analysis and 
 Scientific Computing,  Wiedner Hauptstrasse 8-10, Vienna 1040, Austria.
 }%Lines break automatically or can be forced with \\
\author{Karl Hollaus}%
 \email{karl.hollaus@tuwien.ac.at}
\affiliation{ 
 Technische Universit\"at Wien, Institute for Analysis and Scientific 
 Computing,  Wiedner Hauptstrasse 8-10, Vienna 1040, Austria
}
 %\\This line break forced with \textbackslash\textbackslash
\author{Igor Tsukerman}%
\email{igor@uakron.edu.}
\affiliation{ 
	Department of Electrical and Computer Engineering, The University of Akron, 
	OH 44325-3904, USA
	%\\This line break forced with \textbackslash\textbackslash
}%

\date{\today}% It is always \today, today,

\begin{abstract}
Effective medium theory replaces a given fine-scale heterostructure 
with a homogeneous one in such a way that the physically measurable quantities,
e.g. reaction fields and losses, remain approximately the same.
This Letter shows that the very nature
of the physical problem may change upon homogenization.
A specific example is a stratified
nonlinear magnetic and conducting medium, where a low-frequency excitation
induces eddy currents. It is shown that
the appropriate coarse-scale (homogeneous) model 
is, counter-intuitively, magnetostatic, with an effective 
complex-valued BH-curve whose real and
imaginary parts represent active and reactive losses in the sample.
Similar situations may arise in other physical and engineering
applications -- notably, in diffusion problems with boundary layers.
\end{abstract}

\maketitle

%%%%%%%%%%%%%%%%%%%%%%%%%%%%%%%%%%%%%%%%%%%%%%%%%%%%%%%
\section{Introduction}\label{sec:Intro}
%%%%%%%%%%%%%%%%%%%%%%%%%%%%%%%%%%%%%%%%%%%%%%%%%%%%%%%
%
Effective medium theory (homogenization) is well known 
as an indispensable tool for studying various types of composites 
\cite{Bergman-dielectric-const78,Bensoussan78,Bakhvalov-Panasenko89,Milton02}
and metamaterials 
\cite{Smith06,Bouchitte-Schweizer-homogenization-SRR10,Markel-Schotland12,
	Tsukerman-Markel14,Tsukerman-PLA17}:
a fine-scale heterostructure is replaced with a homogeneous sample
in such a way that its reaction fields or scattering characteristics
are approximately the same \cite{Tsukerman-Markel14}.

In all established homogenization theories, 
the \textit{physical nature of the problem remains unchanged}.
For example, problems involving electric conduction 
on the fine scale are still conduction problems on the coarse
(homogenized) scale, with some effective conductivity defined; 
fine-scale wave problems retain their
physical type upon homogenization, and so on.
It is true that the properties of the effective
medium may in some cases depend strongly on the geometric and physical features
of the microstructure. For example, effective electric conductivity may 
exhibit exponential behavior near the percolation threshold
\cite{Nan-percolation10}; nevertheless this is still a conduction problem
on the coarse scale.

This Letter draws attention to the phenomenon of
``\textit{effective medium transformation}'' -- the
physical and mathematical nature of the problem actually \textit{changing}
upon homogenization. 
One particular case familiar to the authors is that of
stratified magnetic structures composed of thin insulation-coated
conducting magnetic sheets. Such laminated stacks are commonly 
used in magnetic cores to reduce eddy current losses.
On the fine scale, this is an eddy current (EC) problem
(Sections~\ref{sec:curve}, \ref{sec:Formulation}). 
Counterintuitively, upon homogenization
the mathematical and physical model changes from EC to magnetostatic
(Sections~\ref{sec:curve}, \ref{sec:Formulation});
the effective material characteristics are represented by 
\textit{complex $BH$-curves} ($\bfH$ and $\bfB$ are, as usual, the magnetic
field and flux density, respectively). \sectref{sec:Example} 
demonstrates that this homogenized model
yields accurate reaction fields and eddy current losses in the structure.

Layered structures and their homogenization have been extensively studied 
for wave problems (e.g. \cite{Yeh05,Tsukerman-ME-coupling20}
and references there). For fields and eddy currents in laminated
stacks, homogenization has also received much attention
\cite{Gyselinck-Sabariego-Dular06,
	Bottauscio02,Niyonzima13,Niyonzima16,Hollaus15,Hollaus18}
but has never been treated on the coarse scale via complex $BH$ curves.

The fact that the eddy current problem on the fine scale turns into
a magnetostatic one on the coarse scale follows from the analysis
of \sectref{sec:Formulation} (see especially the remark on p.~\pageref{rem:H}).
Loosely speaking, this is because the coarse-scale model can account for
the EC skin effect only indirectly.

%Still, some simplification assumptions are made at this stage
%of development. We use the harmonic balance approximation, well established 
%in the literature in applied and computational electromamgnetics
% \cite{Yamada-harmonic-balance89,
% 	Paoli-Biro-complex-nonlinear98,
%	Ausserhofer-Biro-harmonic-balance07,
%	Biro-nonlinear-periodic14}.

\begin{figure}
	\centering
	\includegraphics[width=0.6\linewidth]{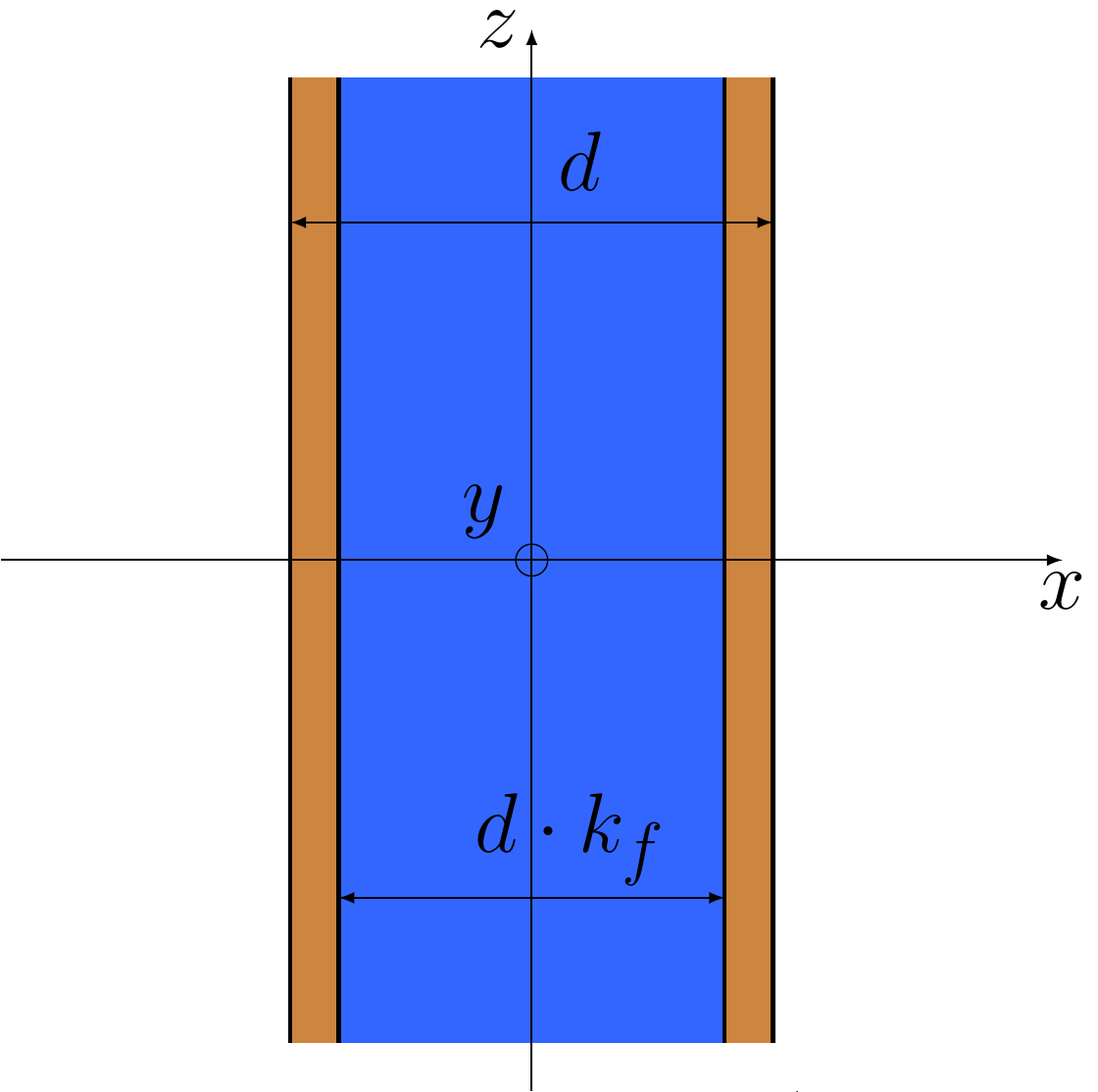}
	\caption{An infinite iron sheet (blue) with   
		insulation (brown).
		The total thickness 
		is $d$, and the thickness of the iron is $k_f d$, 
		$k_f$ being the fill factor. The sheet is assumed 
		to be infinite in the $y$ and $z$ directions.}
	\vskip -4mm
	\label{fig:sheet}
\end{figure}

%%%%%%%%%%%%%%%%%%%%%%%%%%%%%%%%%%%%%%%%%%
\section{A Simplified Model and the Effective BH-Curve}\label{sec:curve}
%%%%%%%%%%%%%%%%%%%%%%%%%%%%%%%%%%%%%%%%%%
%
We start with a simplified setup: a periodic structure
whose lattice cell is an insulated iron sheet, 
infinite in the $y$- and $z$-directions (\figref{fig:sheet});
the geometric parameters are indicated in the figure. 
A constant tangential component $h_0$ of the magnetic field  
$\mathbf{h} = h \hat{z}$ is prescribed at $x = \pm d/2$.
The fine-scale magnetic field $\bfh$ satisfies Maxwell's equations
with the displacement current neglected. 
Under the harmonic balance approximation,
well established in applied electromamgnetic analysis
\cite{Yamada-harmonic-balance89,
	Paoli-Biro-complex-nonlinear98,
	Ausserhofer-Biro-harmonic-balance07,
	Biro-nonlinear-periodic14}, these equations are,
in the SI system under the $\exp(+i \omega t)$ phasor convention,
%
%\begin{equation}\label{eqn:curl-e-curl-h}
$  \nabla \times \bfe = -i \omega \bfb$,
$  \nabla \times \bfh = \bfj$.
%\end{equation}
%
The standard notation $\bfe$, $\bfh$, $\bfb$, $\bfj$
is used for the respective fields and eddy currents;
small letters are reserved for fine-scale quantities;
capital letters will be used later on to indicate coarse-scale 
(homogenized) fields.
The material relations are considered to be intrinsically isotropic
for simplicity:
%
%\begin{equation}\label{eqn:b-eq-muh-j-eq-sigma-e}
$  \bfb \,=\, \mu(|\bfh|) \bfh$;
$  \bfj \,=\, \sigma \bfe$, $\sigma$ being the electric
conductivity.
%\end{equation}
%
For the setup of \figref{fig:sheet}, the eddy current problem 
is one-dimensional: 
\begin{equation}\label{eqn:1D-model}
    -\partial_x \left( \rho \partial_x h(x) \right) \,+\, 
    i \omega \mu (|h(x)|) \, h(x)  \,=\, 0;
    \quad\rho \equiv \sigma^{-1}
\end{equation}  
$$
   \bfh(x) = h(x) \hat{z}; \quad
    h \left(-\frac{d}{2} \right) = h \left(\frac{d}{2}\right) = h_0
$$
%
%\begin{align}
%\mathbf{h}&=\left(\begin{matrix} 0 \\ 0 \\ u(x)\end{matrix}\right), \\
%-\dfrac{\partial}{\partial x}\left( \rho \dfrac{\partial}{\partial x} u\right) 
%+ i\omega\mu (|u|) u &= 0,\\
%u\left(-\frac{d}{2}\right)=u\left(\frac{d}{2}\right)&=\|\mathbf{h}_0\|
%\end{align}
%
The active and reactive losses per unit length
can be written as a single complex number:
\begin{align}
	 P(h_0) \,=\,  \frac{1}{2} \int_{-d/2}^{d/2} \mathbf{e} \mathbf{j}^{*}\, dx  ~+~ 
  \frac{i\omega}{2} \int_{-d/2}^{d/2}  
   \mathbf{h} \, \mathbf{b}^{*} \, dx.
\end{align}
By definition, the coarse-scale field cannot depend on $x$. 
Hence for the infinite sheet problem, the only possible 
effective field is a constant function $\bfH \equiv \bfh_0 = h_0 \hat{z}$. 
Because $\opcurl \bfH = 0$, we define the effective losses as
\begin{align}
\Peff(h_0) \,=\,
\frac{i\omega \mueff(h_0)}{2} 
\int_{-d/2}^{d/2} |h_0|^2 \, dx,
\end{align}
where $\mueff(h_0)$ is a complex number, to be 
determined by demanding $P(h_0) = \Peff(h_0)$. 
Then, by varying $h_0$ and defining 
$\mathbf{B} = \mueff (\bfH) \bfH$, one obtains both 
real and imaginary parts of the effective BH-curve. 
Naturally, this curve depends on the frequency $\omega$, 
the electric conductivity $\sigma$, the thickness 
$d$ and the fill factor $k_f$ of the iron (\figref{fig:sheet}).

Even though the effective complex BH-curve has been defined
for this simplified setup, the numerical example in the
following section shows that this curve gives quite accurate
results for a realistic cylindrical geometry as well
(\figref{fig:lamicores-setup}).
%This suggests that by solving simple 1D problems
%one can derive universal effective BH-characteristics --
%i.e. characteristics independent of the particular geometry of
%the machine under consideration.

%%%%%%%%%%%%%%%%%%%%%%%%%%%%%%%%%%%%%%%%%%%%%%%%%%%%%%
\section{A Realistic Model with Cylindrical Geometry}\label{sec:Formulation}
\begin{figure}
	\centering
	\includegraphics[width=0.85\linewidth]{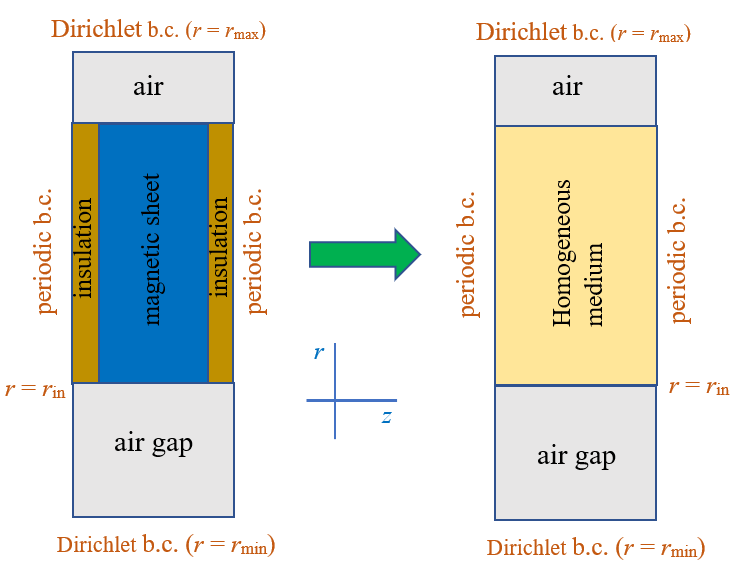}
	\caption{
		Setup: homogenization problem for laminated structures 
		in cylindrical geometry.}
	\label{fig:lamicores-setup}
\end{figure}

%%%%%%%%%%%%%%%%%%%%%%%%%%%%%%%%%%%%%%%%%%%%%%%%%%%%%%
%
We consider cylindrical geometry, typical for rotating machines.
One lattice cell of a periodic stack (\figref{fig:lamicores-setup})
is represented in the cylindrical coordinates 
$(r, z, \theta)$ as a rectangle whose width 
does \textit{not} have to be small relative to the penetration depth
$\delta = (2 \rho / (\omega \mu))^{\frac12}$.
The domain of analysis is 
$\Omegarm = [r_\mathrm{min}, r_\mathrm{max}] \times [-d/2, d/2]$
in the $zr$ plane, 
%Details of the setup can be found in \cite{Schoebinger-Hollaus-Tsukerman20}.
The laminated structure occupies the region 
$[-d/2, d/2] \times [r_\mathrm{in}, r_\mathrm{out}]$, 
of which $-w_{\mathrm{iron}}/2 \leq z \leq 
w_{\mathrm{iron}}/2$ is the iron. 
Periodic boundary conditions link 
the magnetic fields at $z = \pm d/2$. On the exterior boundary
outside the structure ($r = r_{\max} > r_\mathrm{out}$) 
the tangential component of the magnetic field 
is set to zero.
The only conducting material in this model is iron.

%One peculiar feature of the problem is that \textit{eddy currents 
%	and related losses cannot be accounted for on the coarse scale}
%\cite{Niyonzima13,Schoebinger-Hollaus-Tsukerman20}.
%Counter-intuitively, we define a \textit{magnetostatic} rather 
%than an eddy current problem on the coarse scale.

%We assume a $\cos p \theta$ or $\sin p \theta$ variation
%of field and current components in the angular direction 
%($p$ being a given number of pole pairs). This is done to keep
%our focus on the homogenization procedure
%rather than on the technicalities of Fourier analysis,
%and to avoid full-scale 3D simulations of the slot-teeth pattern
%at this stage of development. 

\textit{Remark}.\label{rem:H} The coarse-scale (homogeneous) field $\bfH$
cannot, by definition, depend on $z$ and, from symmetry considerations,
must have a zero $z$ component. 
Hence $\bfJ_{\phi} = (\nabla \times \bfH)_{\phi} = 0$.
Due to insulation, the average $\langle j_z \rangle \equiv J_z = 0$;
then the remaining component $J_r$ is also zero, due to 
$\nabla \cdot \bfJ = 0$.
Hence eddy currents cannot be represented on the coarse scale;
this is why the physical nature of the problem
changes upon homogenization.

For the fine-scale problem, we used the formulation in terms of the 
vector potential $\bft = (t_z, t_r)$ of eddy currents
and the magnetic scalar potential $u$.
Then $\bfh = \bft -\nabla u$;
$\bft$ vanishes outside the conducting regions, as does its tangential
component on the boundary of these regions; also,
$u = 0$ at $r=r_{\max}$. (In general, a Dirichlet boundary condition
	$u = u_D$ is equivalent to $H_\phi = -u_D / r$.)
	
The $\bft$-$u$ formulation (frequently also referred to 
as the $\bfT$-$\Omegarm$ formulation) is very well known
\cite{Carpenter77,Tsukerman90,Schoebinger19,Hollaus18} and has also been
studied in the multiscale setting \cite{Hollaus19}.
The mathematical equations and brief notes on their numerical solution
are included as an Appendix.

On the coarse scale, the effective BH curve is complex-valued 
and derived by equating the
fine- and coarse-scale losses (\sectref{sec:curve}).
There are known instances where complex BH-curves were used in connection with the Preisach hysteresis model\cite{HoBi:02},
but here they are used in the context of homogenization.

The homogenized (coarse-scale) problem is formulated as a magnetostatic one:
\begin{equation}
-\nabla  \cdot \mueff \nabla U \,=\, 0
\end{equation}
with $\bfH = \nabla U$ and the complex magnetic permeability 
$\mueff$. The boundary conditions for $u$ and $U$ are identical.
%
%%%%%%%%%%%%%%%%%%%%%%%%%%%%%%%%%%%%%%%%%%
\section{Numerical Example}\label{sec:Example}
\begin{figure}
	\centering
	\includegraphics[width=0.9\linewidth]{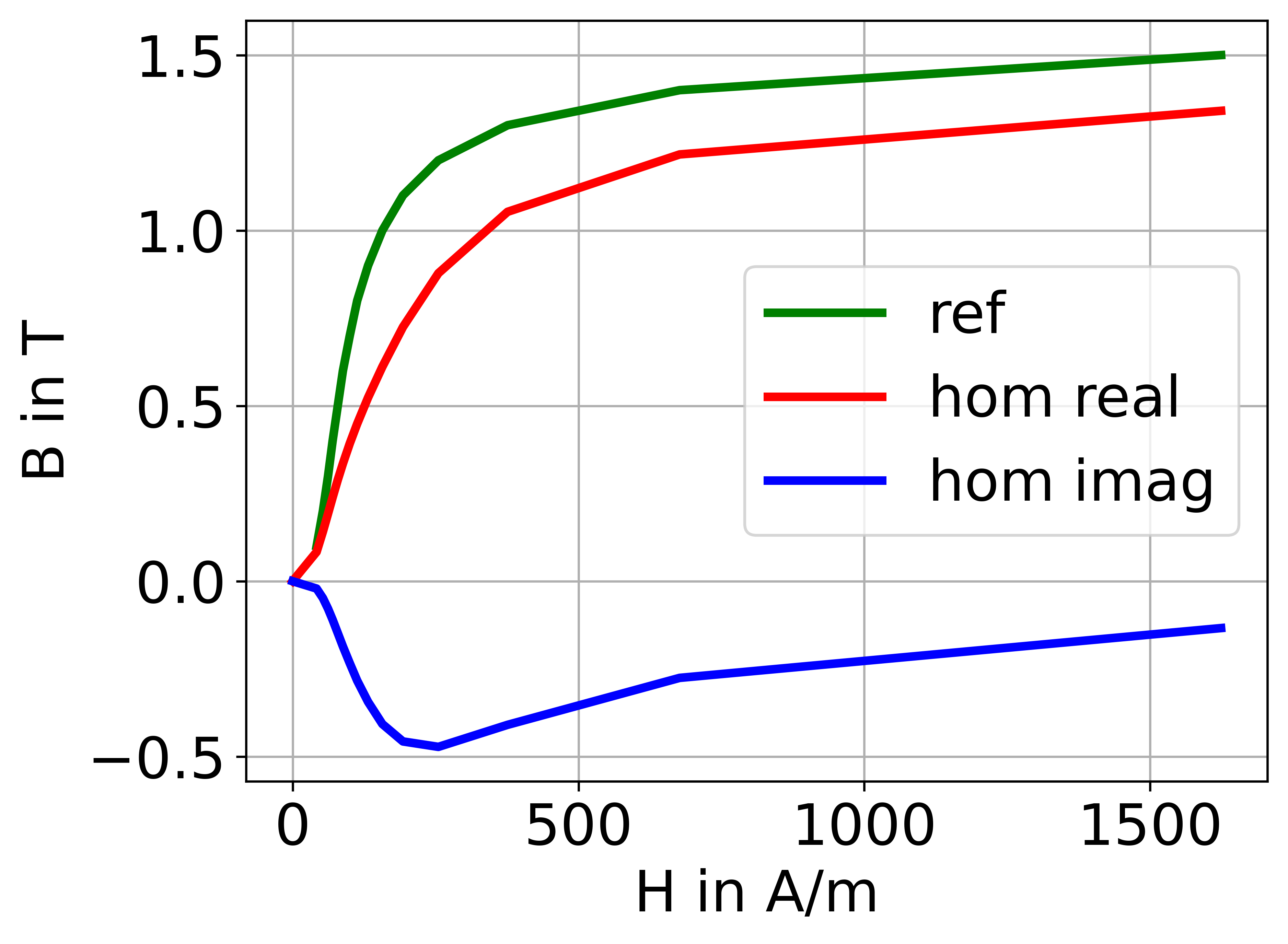}
	\caption{The BH-curve of the example material (green), as well as the real 
	(red) and the imaginary (blue) part of the effective curve at $500$~Hz, 
	obtained by the method described in section \ref{sec:curve}.}
	\label{fig:BHcurve}
\end{figure}

%%%%%%%%%%%%%%%%%%%%%%%%%%%%%%%%%%%%%%%%%%
%
First we consider a cylindrical iron sheet, 
at a frequency of $f = 500$~Hz, with the thickness 
$d=0.5$~mm, the fill factor $k_f=0.9$, and the electric conductivity 
$\sigma = 2\cdot 10^6$~S/m. 
As in the previous section, the effective BH-curve
was determined by equating the fine-scale 
and coarse-scale losses. \figref{fig:BHcurve}
shows the actual and effective BH-curves corresponding to these parameters.
The measurements were taken for a ring core which was incorporated within a computer-aided setup in
acccordance with the international standard IEC 60404-6.

Further, we choose as an example the geometry shown 
in \figref{fig:lamicores-setup}. 
The iron sheet has the inner radius 
of $r_\mathrm{in} = 24$~mm and the outer radius $ r_\mathrm{out} = 30$~mm. 
The inner and outer radii of the air gap are $r_{\min} = 23$~mm 
and $r_{\max}=31$~mm, respectively.

Table \ref{tab:losses} lists the active and reactive losses 
for different ``excitation'' conditions of $u_D$ at $r=r_{\min}$. 
For $u_D \, \lesssim \, 50$~A, the material behaves almost linearly, 
while at $u_D = 300$~A it is nearly fully saturated, as can be seen
by the corresponding values for the $\mathbf{h}$ and $\mathbf{H}$,
evaluated at the bottom of the sheet.
%The influence of the nonlinear material is clearly visible.

\begin{center}
	\begin{table}%[h!]
		\renewcommand{\arraystretch}{1.3}
		\caption{Active Losses $P_a$ and Reactive Losses $P_r$ for Different Exitations at 
		500 Hz}
		\label{tab:losses}
		\centering
		\begin{tabular}{|p{0.6cm}|p{1.2cm}|p{1.2cm}|p{1.2cm}|p{1.2cm}|p{1.2cm}|p{1.2cm}|}
			\hline
			$u_D$ in A & $|\mathbf{h}(r_{in})|$ in $A/m$ & $|\mathbf{H}(r_{in})|$  in $A/m$ & $P_r$, ref. ~in mVA  & $P_r$, hom. in mVA  & 
			$P_a$, ref. in mW  & $P_a$, hom. in mW\\
			\hline\hline
			$50$ & 73.6 & 74.8 & $1.763$ 	 & $1.757$ 	 & $0.7197$ 	 & $0.7643$ 	\\
			$100$ & 115 & 118 & $5.277$ 	 & $5.285$ 	 & $3.001$ 	 & $3.189$ 	\\
			$150$ & 170 & 175 & $11.38$ 	 & $11.46$ 	 & $7.132$ 	 & $7.561$ 	\\
			$200$ & 255 & 264 & $23.22$ 	 & $23.66$ 	 & $12.9$ 	 & $13.59$ 	\\
			$250$ & 443 & 460 & $52.11$ 	 & $53.72$ 	 & $19.01$ 	 & $20.35$ 	\\
			$300$ & 894 & 909 & $122.1$ 	 & $122.2$ 	 & $23.48$ 	 & $25.82$ 	\\
			\hline
		\end{tabular}
	\end{table}
\end{center}

Table \ref{tab:errors} lists the errors of the homogenized solution with 
respect to the actual (fine-scale) one. The error in the potential is the 
difference 
$|U(r)-u(r)|$ at $r=r_{\mathrm{in}}$; this may also serve as
a measure of the difference in the reaction fields.

\begin{center}
	\begin{table}%[h!]
		\renewcommand{\arraystretch}{1.3}
		\caption{Relative Errors for Different Exitations at 500 Hz}
		\label{tab:errors}
		\centering
		\begin{tabular}{|c|c|c|c|}
			\hline
			$u_D$ in A & $\Delta P_r$ in $\%$  & $\Delta P_a$  in $\%$ & 
			$\Delta u(r_{in})$  in $\%$\\
			\hline\hline
			$50$ & $0.33$ 	 & $6.2$ 	 & $2.0$ 	\\
			$100$ & $0.15$ 	 & $6.3$ 	 & $2.7$ 	\\
			$150$ & $0.73$ 	 & $6.0$ 	 & $2.8$ 	\\
			$200$ & $1.9$ 	 & $5.3$ 	 & $2.7$ 	\\
			$250$ & $3.1$ 	 & $7.0$ 	 & $3.6$ 	\\
			$300$ & $0.11$ 	 & $10.0$ 	 & $1.7$ 	\\
			\hline
		\end{tabular}
	\end{table}
\end{center}

\begin{figure}
	\centering
	\includegraphics[width=0.99\linewidth]{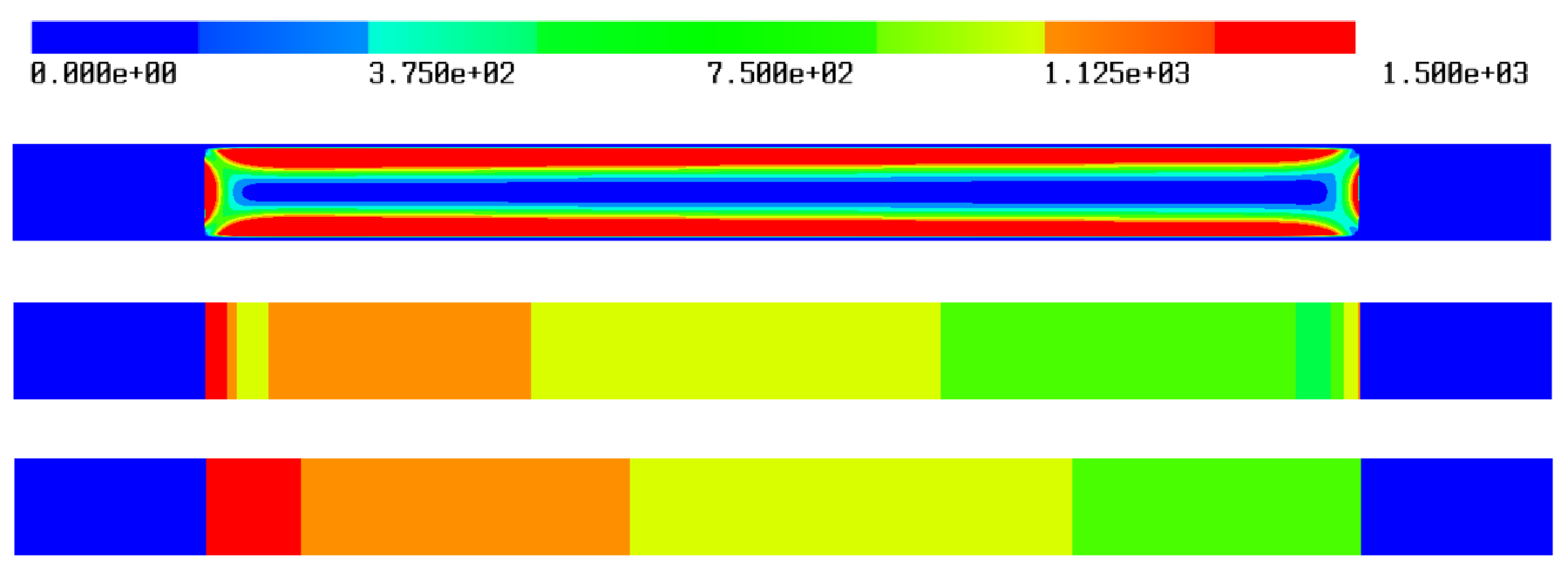}
	\caption{The eddy current losses in $W/\mathrm{m}^3$ for the reference 
		solution 
		(top), the reference solution averaged over the thickness (middle) 
		and the homogeneous solution (bottom) for the excitation 
		$u_D = 100$~A at $500$~Hz.}
	\label{fig:losses1}
\end{figure}

It can be seen that all errors stay in the range of a few percent with a slight 
increase for higher saturation. Note that an error of zero would be impossible 
even for a fully linear material, because the effective material parameters 
were obtained from the infinite sheet and therefore cannot fully
account for the edge effects, which are most prevalent at the shorter
sides of the sheet.

\begin{figure}
	\centering
	\includegraphics[width=0.9\linewidth]{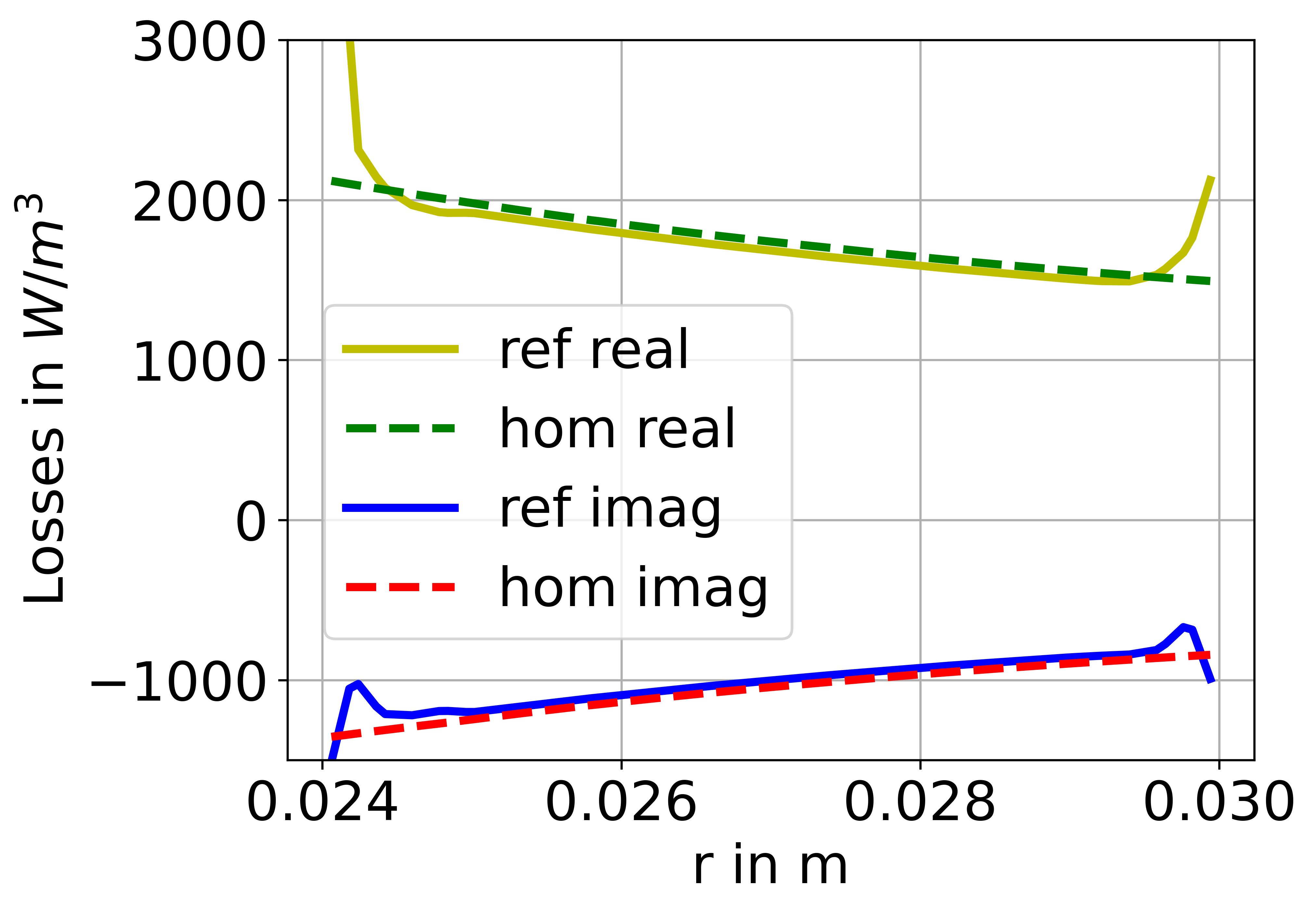}
	\caption{The complex losses for the fine-scale solution (solid lines, 
	averaged over the thickness) and the homogeneous solution (dashed lines) 
	along the midline $x=0$ for the excitation $u_D=100$~A at $500$~Hz.}
	\label{fig:losses2}
\end{figure}

\begin{figure}
	\centering
	\includegraphics[width=0.9\linewidth]{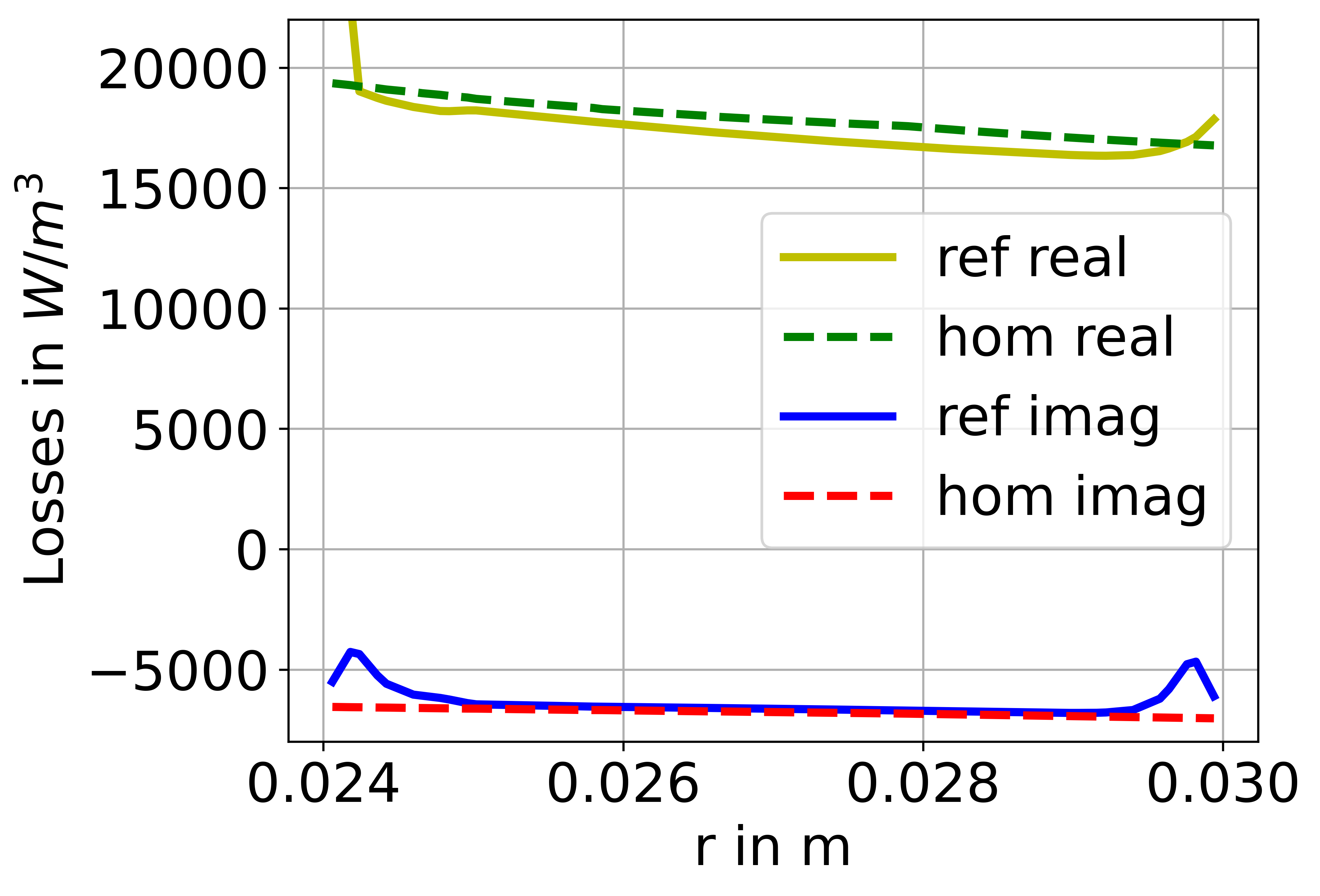}
	\caption{The complex losses for the fine-scale solution (solid lines, 
	averaged over the thickness) and the homogeneous solution (dashed lines) 
	along the midline $x=0$ for the excitation $u_D=250$~A at $500$~Hz.}
	\label{fig:losses3}
\end{figure}

Comparison of the distributions of the loss densities on the fine
and coarse scales demonstrates that the homogenized problem accurately 
represents not only global but also local losses (\figref{fig:losses1}).
\figref{fig:losses2} shows the active and reactive components 
of the losses along the midline $x=0$, at low saturation of the iron. 
\figref{fig:losses3} shows similar distributions at a higher 
saturation. It can be seen that the homogeneous solution reproduces 
the distribution accurately, except for the edge effects, at low and high 
saturations.

\begin{figure}
	\centering
	\includegraphics[width=0.9\linewidth]{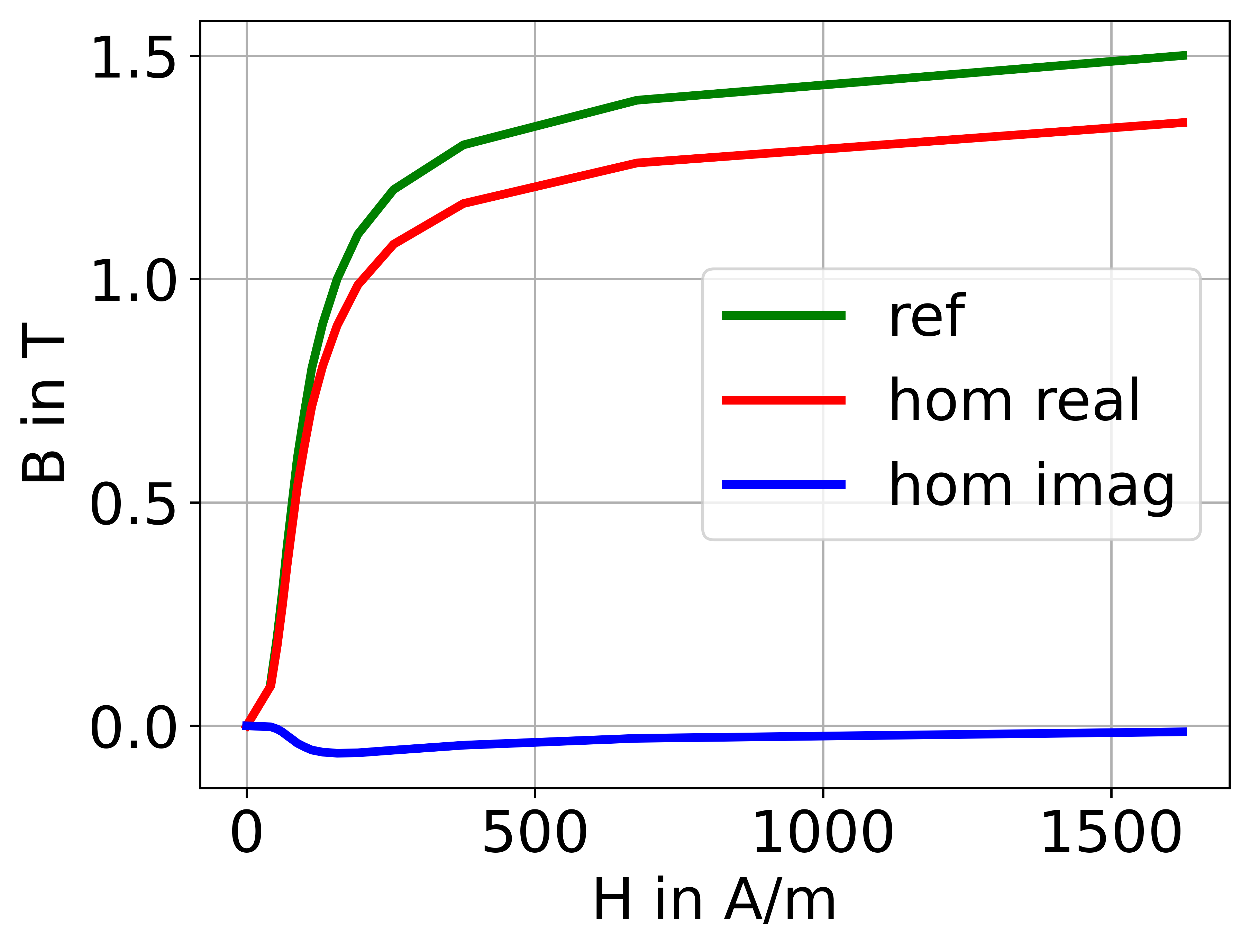}
	\caption{The BH-curve of the example material (green), as well as the real 
		(red) and the imaginary (blue) parts of the effective curve at $f = 
		50$~Hz.}
	\label{fig:BHcurve50}
\end{figure}

For comparison, the BH-curve for the same material but at $f = 50$~Hz
are displayed in \figref{fig:BHcurve50}.
As can be seen in Table \ref{tab:errors50}, the results are lower 
overall than at $f = 500$~Hz.

\begin{center}
	\begin{table}[h!]
		\renewcommand{\arraystretch}{1.3}
		\caption{Relative Errors for Different Exitations at 50 Hz}
		\label{tab:errors50}
		\centering
		\begin{tabular}{|c|c|c|c|}
			\hline
		$u_D$ in A & $\Delta P_r$ in $\%$  & $\Delta P_a$  in $\%$ & 
			$\Delta u(r_{in})$  in $\%$\\
			\hline\hline
			$50$ & $0.07$ 	 & $4.8$ 	 & $0.3$ 	\\
			$100$ & $0.4$ 	 & $5.6$ 	 & $0.34$ 	\\
			$150$ & $0.16$ 	 & $4.9$ 	 & $0.32$ 	\\
			$200$ & $0.6$ 	 & $5.3$ 	 & $0.61$ 	\\
			$250$ & $1.2$ 	 & $7.3$ 	 & $0.97$ 	\\
			$300$ & $3.0$ 	 & $11.2$ 	 & $2.4$ 	\\
			\hline
		\end{tabular}
	\end{table}
\end{center}

%%%%%%%%%%%%%%%%%%%%%%%%%%%%%%%%%%%%%%%%%%
\section{Conclusion}
%%%%%%%%%%%%%%%%%%%%%%%%%%%%%%%%%%%%%%%%%%
%
The fine-scale  eddy current problem examined in this Letter 
changes its physical nature upon homogenization, when a stratified
nonlinear magnetic medium is replaced with an effective homogeneous
one. Namely, the appropriate coarse-scale (homogeneous) model 
is, counter-intuitively, magnetostatic, with a nonlinear 
complex-valued BH-curve whose real and
imaginary parts represent active and reactive losses in the iron.
In addition, the reaction field of the structure
is also represented accurately on the coarse scale.
These results may be relevant in other physical and engineering
applications, where similar problems -- especially diffusion
problems with boundary layers -- may arise.

%%%%%%%%%%%%%%%%%%%%%%%%%%%%%%%%%%%%%%%%%%%%%%%%%%%%%%
\section*{Appendix: The Mathematical Formulation}\label{app:Math-formulations}
%%%%%%%%%%%%%%%%%%%%%%%%%%%%%%%%%%%%%%%%%%%%%%%%%%%%%%
%
The key equations for the eddy current problem
under consideration are, in weak form,
\begin{equation}\label{eqn:gradu-gradu-T}
   (\mu \nabla u, \nabla u') - (\mu \bft, \nabla u') \,=\,
   -(\mu \nabla u_D, \nabla u')
\end{equation}
%
%\vskip -8mm
\begin{equation}\label{eqn:curl-curl-T}
   (\sigma^{-1} \nabla \times \bft, \nabla \times \bft') 
   - i \omega (\mu \bft, \bft') + i \omega (\mu \nabla u, \bft')
    \,=\, 0.
\end{equation}
In these equations, parentheses denote the $L_2^3$ inner products
of vector functions; 
$u, u' \in H_0^1(\Omegarm)$, 
$\bft, \bft' \in H_0(\mathrm{curl}, \Omegarm_{\mathrm{iron}})$;
the primed symbols denote arbitrary test functions 
in their respective spaces.
Definitions of the function spaces and their discretizations
using high-order edge elements for $\bft$ and nodal elements for $u$
 can be found in \cite{SchoeZagl:05}. All simulations were done using the 
open-source package Netgen/NGSolve \cite{NGSolve}.

%%%%%%%%%%%%%%%%%%%%%%%%%%%%%%%%%%%%%%%%%%
\section*{Acknowledgment}
%%%%%%%%%%%%%%%%%%%%%%%%%%%%%%%%%%%%%%%%%%
%
This work was supported in part by the Austrian Science Fund (FWF) under Project P 31926.
 Research of IT was supported in part by the US National Science Foundation awards DMS-1216970 and
DMS-1620112.

\bibliographystyle{unsrt}

\nocite{*}
\bibliography{effective_medium_transformation}% Produces the bibliography via 
%BibTeX.

\end{document}